\documentclass[conference]{IEEEtran}
\IEEEoverridecommandlockouts

\usepackage{multirow}
\usepackage{pifont}
\usepackage{stfloats}
\usepackage[table]{xcolor}
\usepackage{booktabs}
\usepackage{array}  
\usepackage{cite}
\usepackage{amsmath,amssymb,amsfonts}
\usepackage{algorithmic}
\usepackage{graphicx}
\usepackage{textcomp}
\usepackage{xcolor}
\usepackage{url}  
\usepackage[hidelinks]{hyperref}  
\usepackage{svg}
\def\BibTeX{{\rm B\kern-.05em{\sc i\kern-.025em b}\kern-.08em
    T\kern-.1667em\lower.7ex\hbox{E}\kern-.125emX}}
\begin{document}

\title{Understanding Everything as Code: A Taxonomy and Conceptual Model}
\author{\IEEEauthorblockN{1\textsuperscript{st} Haoran Wei}
\IEEEauthorblockA{\textit{Western University} \\
London, Canada \\
hwei53@uwo.ca}
\and
\IEEEauthorblockN{2\textsuperscript{nd} Nazim Madhavji}
\IEEEauthorblockA{\textit{Western University} \\
London, Canada \\
nmadhavji@uwo.ca}
\and
\IEEEauthorblockN{3\textsuperscript{rd} John Steinbacher}
\IEEEauthorblockA{\textit{IBM Canada Lab} \\
Toronto, Canada \\
jstein@ca.ibm.com}
}
\maketitle
\begin{abstract}
Background: Everything as Code (EaC) is an emerging paradigm aiming to codify all aspects of modern software systems. Despite its growing popularity, comprehensive industry standards and peer-reviewed research clarifying its scope and guiding its adoption remain scarce. Aims: This study systematically analyzes existing knowledge and perceptions of EaC, clarifies its scope and boundaries, and provides structured guidance for researchers and practitioners. Method: We conducted a large-scale multivocal literature review (MLR), synthesizing academic and grey literature sources. Findings were analyzed quantitatively and thematically. Based on this analysis, we developed a taxonomy and conceptual model of EaC, validated through collaboration with industry experts. Results: The resulting taxonomy comprises 25 distinct EaC practices organized into six layers based on industry awareness and functional roles. The conceptual model illustrates focus areas, overlaps, and interactions among these EaC practices within the software delivery lifecycle. Additionally, practical code examples demonstrating the implementation of these practices were developed in collaboration with industry experts. Conclusions: This work addresses the current scarcity of academic discourse on EaC by providing the first comprehensive taxonomy and conceptual model. These contributions enhance conceptual clarity, offer actionable guidance to practitioners, and lay the groundwork for future research in this emerging domain.
\end{abstract}


\section{Introduction}
In recent years, Infrastructure as Code (IaC) has become indispensable for efficiently managing and provisioning computing environments through machine-readable definitions \cite{i1}. IaC primarily addresses infrastructure components such as servers, networks, and storage, yet other elements of the software delivery lifecycle (such as application configurations and operational pipelines) remain beyond its traditional scope \cite{i2}. Building upon the proven effectiveness of IaC, the Everything as Code (EaC) paradigm expands this foundational principle by advocating for all operational aspects—including infrastructure, configurations, security policies, and Continuous Integration/Continuous Delivery (CI/CD) pipelines—to be represented and managed uniformly as code \cite{i3}. This unified, code-centric approach aims to enhance automation, consistency, and traceability throughout the software delivery process \cite{i4}.

Given these promising benefits, EaC is gaining substantial traction and is increasingly recognized as the future direction of IT operations \cite{i5,i6,i7}. However, the adoption of EaC presents notable challenges, primarily due to the requisite shift in organizational mindset. Unlike IaC, EaC necessitates the codification of a much broader range of processes and resources, often surpassing the capabilities of existing DevOps practices \cite{i8}. Consequently, clear industry standards and comprehensive scientific literature are needed to provide empirical evidence and structured guidance for EaC adoption. Currently, there is no dedicated standard—comparable to ISO/IEC/IEEE 32675 for DevOps—that provides explicit, comprehensive guidelines for EaC. Furthermore, our extensive literature search indicates a notable scarcity of peer-reviewed academic research on EaC (see Fig. \ref{fig:search_results}). To the best of our knowledge, no comprehensive empirical studies on EaC currently exist (detailed further in Section II).

Due to this gap, developers are often compelled to rely on non-peer-reviewed resources such as blogs and online forums. However, these informal resources typically provide generic advice or subjective opinions, rather than rigorously validated guidelines \cite{i9}. Our preliminary analysis highlights a significant lack of consensus among these sources regarding the standardized scope, boundaries, and interactions of EaC practices, complicating informed decision-making for practitioners (detailed in Section II).

To address these gaps, we conducted a large-scale multivocal literature review \cite{m1} on EaC, synthesizing findings through both quantitative and thematic analysis \cite{m2}. Based on this comprehensive review, we propose a detailed taxonomy to clarify the scope of EaC. Our taxonomy identifies 25 distinct EaC practices (e.g., Infrastructure as Code, Configuration as Code, and Policy as Code) and classifies them according to their industry awareness and functional roles. Complementing this taxonomy, we developed a conceptual model that illustrates the focus areas, overlaps, and interrelationships among the various EaC practices across the software delivery lifecycle.

Furthermore, we collaborated closely with industry experts to validate and refine both our taxonomy and conceptual model, ensuring their accuracy, relevance, and applicability (validation details are provided in Section VI). 

The taxonomy (Section IV) and conceptual model (Section V) constitute the primary contributions of this paper. Additionally, in collaboration with industry practitioners, we have developed practical code examples to demonstrate the implementation of our conceptual model using industry-standard tools. These examples serve as supplementary contributions. 

Collectively, these contributions aim to provide improved conceptual clarity, actionable guidance for practitioners adopting EaC practices, and a foundation for future academic research in this domain.

\section{Related Work and Motivation}\label{sec:related_work}
\begin{figure}[b]
\centerline{\includegraphics[width=0.8\columnwidth,height=\textheight,keepaspectratio]{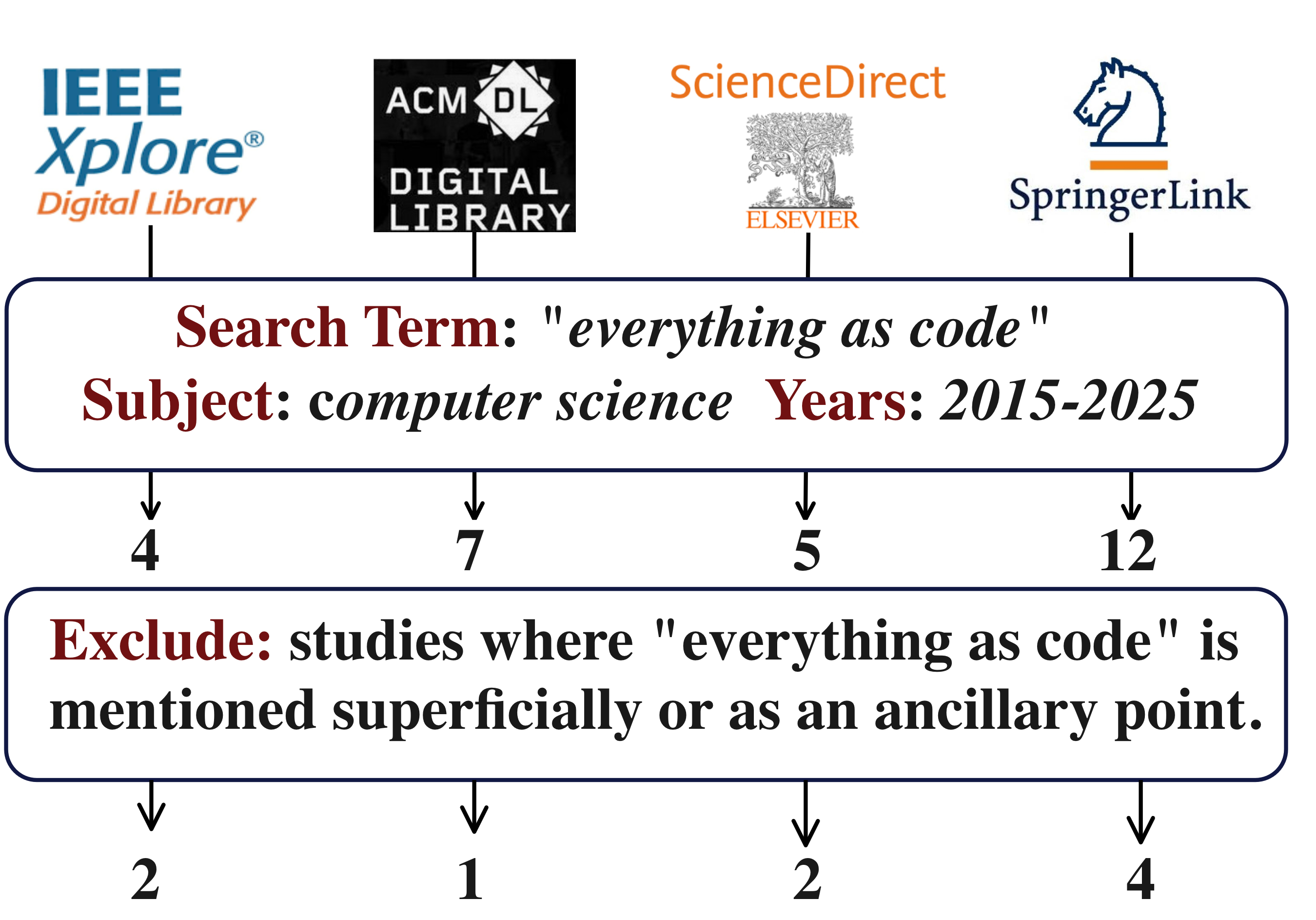}}
\caption{Numbers of publications on “Everything as Code” in major academic databases, searched in March 2025.}
\label{fig:search_results}
\end{figure}

Despite the growing significance of Everything as Code (EaC) in modern software delivery~\cite{i5}, peer-reviewed literature explicitly addressing it remains scarce, as shown in Fig.~\ref{fig:search_results}. Existing studies are limited in scope or depth. Stirbu et al.~\cite{i3} identified Infrastructure as Code (IaC) and Configuration as Code (CaC)~\cite{b9} as EaC subsets but focused on non-software contexts (e.g., healthcare compliance), limiting applicability to software engineering. Yanagawa et al.~\cite{b2} proposed a framework combining Policy as Code (PaC)~\cite{b10} and Compliance as Code (CoaC)~\cite{b11}, narrowly defined EaC through this integration. Haverinen et al.~\cite{b3} explored automation for compliance and security, addressing a subset of EaC. Zhao et al.~\cite{b4} summarized DevSecOps tools, briefly mentioning Security as Code (SaC)~\cite{b13}, PaC, and CoaC. Several book chapters \cite{b5,b6,b7} also touched upon EaC's implications within software delivery, albeit superficially.

A comprehensive empirical study that synthesizes EaC’s scope, key as-code practices, and their interrelationships is absent. Current work either focuses on specific practices or lacks rigorous data-driven analysis.

\begin{figure}[t]
\centerline{\includegraphics[width=1.02\columnwidth,height=\textheight,keepaspectratio]{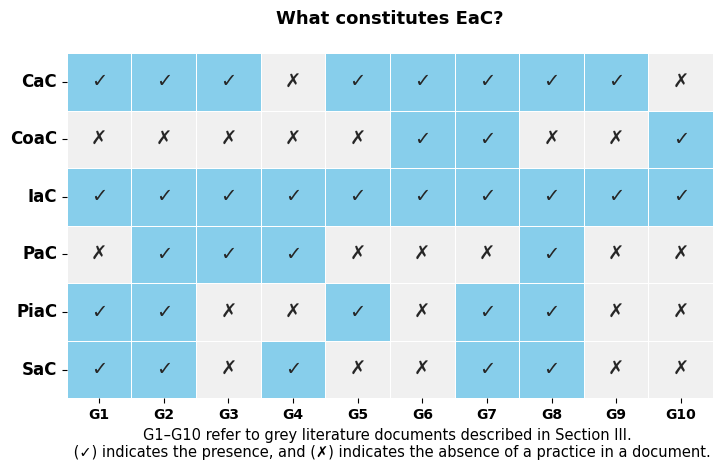}}
\caption{Infrastructure as Code (IaC) and Configuration as Code (CaC) consistently emerged as foundational practices within EaC. Other as-code practices, including Policy as Code (PaC), Compliance as Code (CoaC), Pipeline as Code (PiaC), and Security as Code (SaC), are frequently, but inconsistently, identified as elements of EaC.}
\label{fig:diverge}
\end{figure}

Given this limitation, we expanded our investigation to include grey literature (e.g., white papers, industry reports, and blogs). Fig.~\ref{fig:diverge} illustrates divergent perspectives regarding key EaC components identified in the grey literature sources.  The varied interpretations captured in Fig.~\ref{fig:diverge} highlight the absence of consensus among practitioners regarding the standardized scope of EaC.

Additionally, certain as-code practices—particularly PaC and CoaC—display overlapping functionalities and tightly coupled interdependencies, lacking clear boundaries or a well-articulated separation of concerns. For instance, Yanagawa et al. \cite{b2} characterized CoaC primarily as expressing regulatory controls as code, positioning PaC as a downstream mechanism for their validation and enforcement. Whereas, some industry practitioners conceptualize CoaC as a subset of PaC dedicated exclusively to enforcing regulatory compliance \cite{d14,d15}. Conversely, others adopt a broader definition, applying CoaC to describe comprehensive, programmatic methods for automating the entire compliance lifecycle—encompassing implementation, validation, remediation, monitoring, and reporting—thus surpassing PaC’s scope \cite{d10,d11,d12,d13}.

Furthermore, existing practitioner discourse largely addresses EaC at a conceptual or motivational level rather than providing detailed, actionable implementation guidance. Consequently, developers may find the topic overwhelming and ambiguous, limiting practical adoption.

From our investigation, we have identified the following gaps in the current state of EaC research and practice:
\begin{itemize}
\item \textbf{Scarcity of academic literature}: A pronounced lack of academic contributions limits comprehensive understanding and systematic exploration of EaC.
\item \textbf{Ambiguity and lack of consensus}: There is ambiguity and disagreement regarding the scope, boundaries, and interplay among various EaC components.
\item \textbf{Limited practical guidance}: Conceptual ideals underpinning EaC have yet to be translated effectively into actionable, implementable strategies.
\end{itemize}
These identified gaps motivate this study.

\section{Foundational Knowledge Base}\label{sec:MLR}
This section describes our methodology for assembling the foundational knowledge required to construct a taxonomy and conceptual model of Everything as Code (EaC). Specifically, we conducted a multivocal literature review (MLR), following established guidelines in software engineering research \cite{m1}, to select and evaluate relevant literature.
\subsection{Literature Collection Strategy and Process}

\begin{figure}[b] \centerline{\includegraphics[width=\columnwidth,height=\textheight,keepaspectratio]{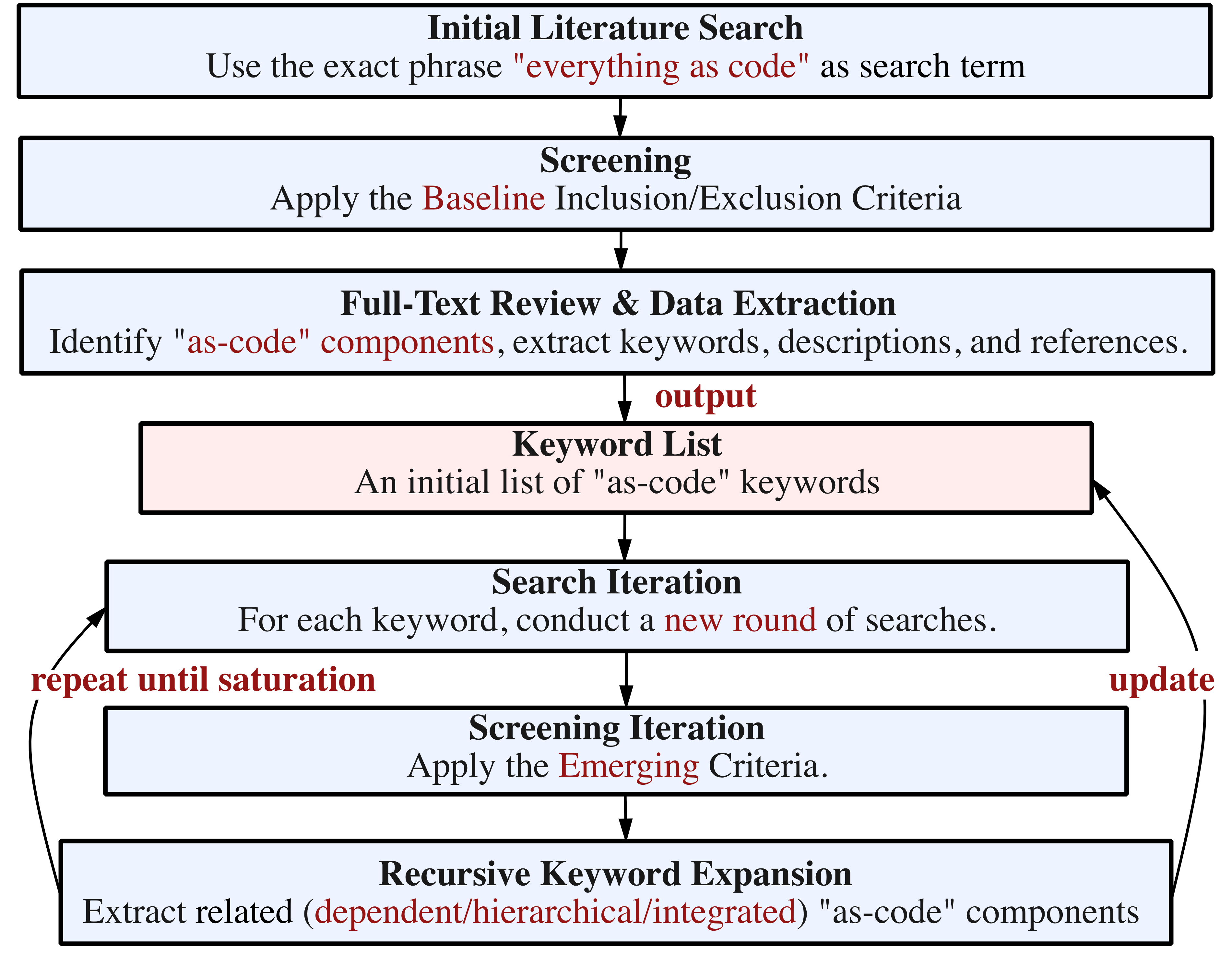}} \caption{Literature Collection Process.} \label{fig:collection} \end{figure}

Fig.~\ref{fig:collection} illustrates our literature-collection process. Scientific studies were gathered from major academic databases—IEEE Xplore, ACM Digital Library, ScienceDirect, and SpringerLink. Grey literature documents (e.g., white papers, reports, and practitioner blogs) were collected through Google searches, prioritizing sources from leading cloud providers (e.g., AWS, Azure, and Google Cloud), DevOps vendors (e.g., HashiCorp, GitLab, and Red Hat), and technical blogging platforms (e.g., Medium, LinkedIn, and The New Stack).

We applied the baseline inclusion and exclusion criteria, outlined in Table~\ref{tab:criteria}, to filter relevant documents. After document selection, we performed a thorough full-text review to identify and catalog distinct as-code practices explicitly mentioned or implicitly indicated in the literature (e.g., IaC, CaC, PaC). This step provided an initial set of keywords reflecting key EaC components.

Using these identified keywords, we iteratively conducted additional literature searches to uncover further as-code practices and related literature. During this process, we continuously applied the emerging criteria in Table~\ref{tab:criteria}. This recursive, keyword-expansion strategy was crucial for comprehensively capturing the breadth of EaC practices. 

We collected and evaluated a total of 286 literature documents; our final selection (after screening) includes 42 peer-reviewed scientific papers and 86 grey literature documents, available in Section~\ref{sec:resources}). For reference clarity, we designate these sources as S1–S42 (scientific literature) and G1–G86 (grey literature) throughout this paper. This corpus serves as the knowledge base for constructing the taxonomy and conceptual model of EaC.

\begin{table}[t]
\caption{Inclusion and Exclusion Criteria}
\begin{tabular*}{\columnwidth}{@{\extracolsep{\fill}}lp{0.82\columnwidth}}
\toprule
\textbf{Inclusion} & \textbf{(Baseline)} Sources that explicitly discuss EaC or multiple distinct ``as-code'' paradigms. \\
\cmidrule(l){2-2}
& \textbf{(Baseline)} Documents that provide detailed discussions on the scope, definition, or conceptual frameworks of EaC. \\
\cmidrule(l){2-2}
& \textbf{(Emerging)} Documents that describe a specific new as-code practice and demonstrate how it complements EaC. \\
\midrule
\textbf{Exclusion} & \textbf{(Baseline)} Brief or superficial mentions of EaC without substantial insights into its scope or components. \\
\cmidrule(l){2-2}
& \textbf{(Baseline)} Vendor-specific content focused on product promotion rather than generalizable EaC concepts or frameworks. \\
\cmidrule(l){2-2}
& \textbf{(Emerging)} Documents that mention new as-code practices only in passing, without substantial details or context that adds to the understanding of its role in EaC.\\
\bottomrule
\end{tabular*}
\label{tab:criteria}
\end{table}

\begin{table}[t]
    \centering
    \caption{Data Extraction Form}
    \label{tab:data-extraction-form}
    \begin{tabular*}{\columnwidth}{@{\extracolsep{\fill}}p{0.22\columnwidth}p{0.7\columnwidth}}
        \toprule
        \textbf{Data Item} & \textbf{Description} \\
        \midrule
        Metadata & Literature ID, title, authors, year, and source type. \\
        \midrule
        EaC Scope & Identify and list all explicitly mentioned or implicitly indicated as-code components/practices. \\
        \midrule
        Definitions & Summarize how the authors define each identified EaC component/practice. \\
        \midrule
        Inter-component Relationships & Describe relationships, overlaps, dependencies, or integrations between different EaC components as discussed in the literature. \\
        \midrule
        Ambiguities/Gaps & Identify and document ambiguities, open questions, or gaps explicitly stated or implied by the authors regarding EaC concepts. \\
        \midrule
        Implementation Guidance & Summarize practical recommendations, tools, or guidelines provided for implementing EaC practices. \\
        \bottomrule
    \end{tabular*}
    \label{tab:extraction}
\end{table}

\subsection{Data Extraction}
To systematically extract and synthesize information required to build the EaC taxonomy and conceptual model, we designed a structured data extraction form (Table~\ref{tab:extraction}). The extraction was initially carried out by the first author. Subsequently, the second author performed a detailed validation by tracing extracted data back to original literature sources, ensuring the accuracy of captured information. The extracted data is available in the supplementary resources (Section \ref{sec:resources}).

\section{A Taxonomy of EaC Practices}
\begin{table*}[t]
\centering
\caption{Overview of the Taxonomy Design}
\label{tab:taxo_design}
\begin{tabular}{@{}lll@{}}
\toprule
\multicolumn{3}{c}{\textbf{Objects}: 25 As-Code Practices in Everything as Code} \\
\midrule
\multicolumn{3}{c}{\textbf{Meta-Characteristic}: Recognition and Utilization of `as Code' Practices in Software Development} \\
\midrule
\textbf{Dimensions} & \textbf{Industry Awareness and Tooling Support} & \textbf{Functionality and Application} \\
\midrule
\textbf{Method} & Conceptual-to-empirical (top-down) & Empirical-to-conceptual (bottom-up) \\
\midrule
\textbf{Characteristics (Measures)} & Literature Frequency, Tooling Availability & Functional Role, Implementation Area \\
\midrule
\textbf{Categories} & 
\begin{tabular}[c]{@{}l@{}}
Established Practices \\
Emerging Practices \\
\end{tabular} & 
\begin{tabular}[c]{@{}l@{}}
Infrastructure Provisioning and Management\\
Platform and Orchestration\\
Application Design and Development\\
Data and Database\\
Security and Compliance\\
Observability and Analysis\\
\end{tabular} \\
\midrule
\textbf{Ending Conditions} & 
\multicolumn{2}{l}{
\begin{tabular}[c]{@{}l@{}}
1) All objects have been examined. 2) No object was merged with a similar object in the last iteration. \\
3) At least one object is classified under every category of every dimension.
\end{tabular}} \\
\bottomrule
\end{tabular}
\end{table*}

This section describes our methodological approach for developing a multi-dimensional taxonomy of EaC practices and presents the taxonomy. We employed the taxonomy development method proposed by Nickerson et al. \cite{m3}, as updated in~\cite{m4}. Table \ref{tab:taxo_design} summarizes the taxonomy design.

Our primary objective is to clearly define and structure the scope of EaC practices, highlighting their key and complementary components through a hierarchical organization. Central to this investigation is the extent of recognition of these practices within the industry, their categorization into key versus complementary roles, and their practical applicability in software development and operations contexts.

To achieve this, we established two dimensions for our taxonomy: (1) \emph{Industry Awareness and Tooling Support}, designed to reflect the level of industry recognition, adoption, and available tooling; and (2) \emph{Functionality and Application}, intended to categorize EaC practices based on their operational functionality and specific implementation contexts.

\subsection{Dimension 1: Industry Awareness and Tooling Support}
We adopted the Conceptual-to-Empirical (top-down) approach~\cite{m3} for developing this dimension. Starting from the 25 EaC practices identified in our literature set, we analyzed two complementary indicators: (1) \emph{Literature Frequency}: the proportion of the reviewed sources that explicitly discuss a practice in the context of EaC; and (2) \emph{Tooling Availability}: the availability of publicly documented tools that implement the practice.

\begin{figure}[b]
\centerline{\includegraphics[width=\columnwidth,height=\textheight,keepaspectratio]{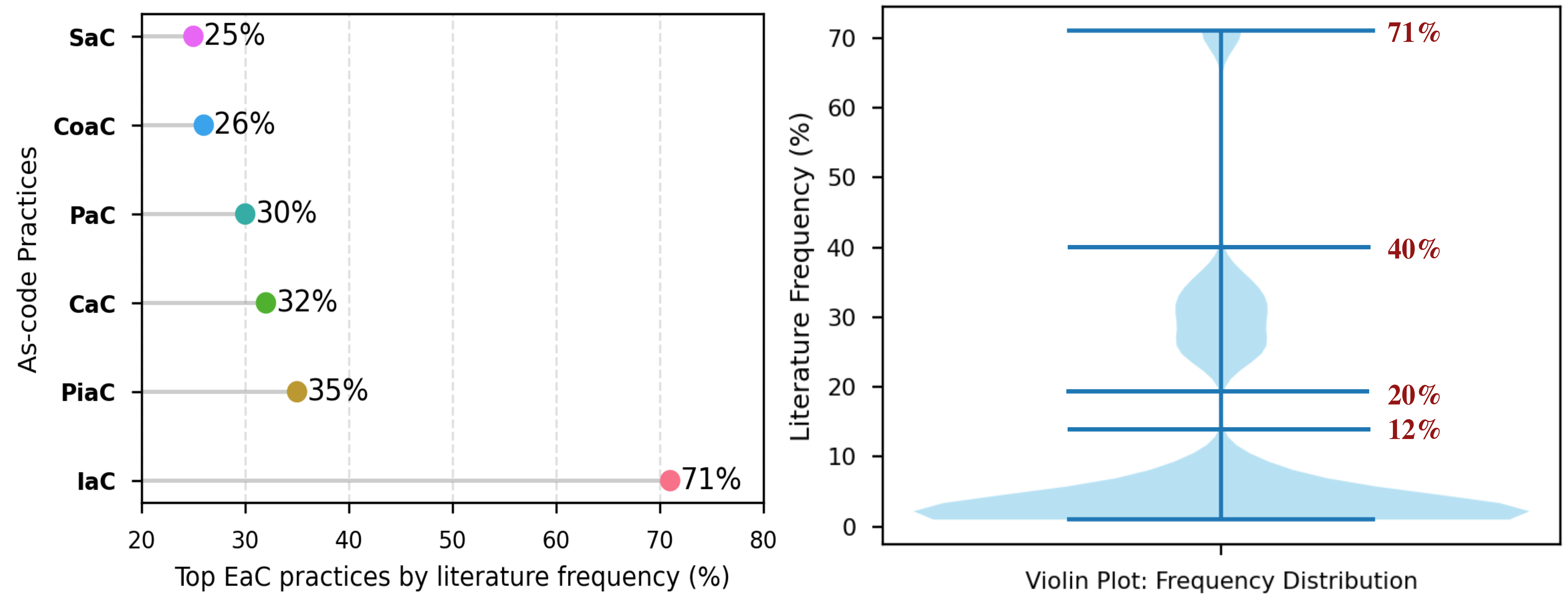}}
\caption{Frequencies and frequency groupings of EaC practices.}
\label{fig:violin}
\end{figure}
Infrastructure as Code (IaC) was referenced in 91 sources (71\%), confirming its centrality to EaC. The remaining practices clustered into a noticeably lower‑frequency band, yielding the trimodal distribution illustrated in the violin plot~\cite{m5} in Fig.~\ref{fig:violin}. Because IaC dominates the first peak, we treat it as an outlier when forming categories. Practices that appear frequently in the literature also exhibit richer ecosystems of supporting tools, suggesting a correlation between academic/industrial attention and practical adoption. Combining both indicators, we created two classes: 1) \textbf{Established Practices}:  well-recognized in the community, and supported by a comprehensive array of tools. 2) \textbf{Emerging Practices}: gaining traction, evidenced by some documents and emerging tooling support. Specifically, a practice is classified as an “Established Practice” if its literature frequency falls within the upper frequency band (20\% - 40\%) and there are more than three documented tools for that practice in the marketplace, according to the extracted data from the collected literature. Otherwise, it is categorized as an “Emerging Practice”. Table~\ref{tab:practices} summarizes this dimension of our taxonomy.

\begin{table*}[t!]
\centering \caption{The Taxonomy of EaC practices: Industry Awareness and Tooling Support}
\label{tab:practices}
\begin{tabular*}{\textwidth}{@{\extracolsep{\fill}}p{0.14\textwidth}p{0.45\textwidth}p{0.22\textwidth}p{0.11\textwidth}@{\extracolsep{\fill}}}
\toprule
\textbf{Practice} & \textbf{Description} & \textbf{Tools} & \textbf{References} \\
\midrule
\multicolumn{4}{c}{\cellcolor{gray!20}\textbf{Established Practices:} Practices that are well-recognized, frequently discussed in the community, and supported by a comprehensive array of tools.} \\
\midrule
Infrastructure as Code & Manage and provision IT infrastructure (e.g., compute, storage, and network) through code. & Terraform, Pulumi, CloudFormation, Ansible, Chef, Puppet & S1-15, G55-57 \\
Configuration as Code & Manage post‑provision configuration (operating system, middleware, and application settings) through code. & Ansible, Chef, Puppet, SaltStack, Helm & S16-19, G23-31 \\
Pipeline as Code & Define and automate workflows, such as CI/CD, infrastructure, or data processing, through version-controlled code  & Jenkins, GitLab CI, GitHub Actions, CircleCI, Argo CD & S25-28, G44-45 \\
Policy as Code & Express governance, security, or business rules in machine‑readable form so policy engines can validate and enforce them. & Open Policy Agent (OPA), Sentinel, Azure Policy, Kyverno & S29-31, G32-43 \\
Security as Code & Define and automate security controls, tests, and scans through code within DevOps pipelines. & Checkov, Trivy, Snyk, SonarQube, Gauntlt & S32-38, G46-47 \\
Compliance as Code & Encode regulatory controls and evidence as executable artefacts, automating assessment, remediation, and reporting of compliance posture. & OSCAL-COMPASS, Trestle, Auditree, InSpec, OPA & S20-24, G48-54 \\
\midrule
\multicolumn{4}{c}{\cellcolor{gray!20}\textbf{Emerging Practices:} Practices that are gaining traction, evidenced by some documented effectiveness and emerging tool support.} \\
\midrule
SLO as Code & Define Service Level Objectives in declarative formats. & OpenSLO & S39-42, G85-86 \\
Detection as Code & Use code to manage security detection rules and response processes. & Datadog & G66-67 \\
Monitoring as Code & Define monitoring configurations (logs, metrics, and alerts) in code for consistent observability. & Nagios, Zabbix, Datadog & G69-70 \\
Dashboards as Code & Version and automate the creation of metrics dashboards. & Grafana & G70-71 \\
Analytics as Code & Automate analytics setup and user behavior tracking via code. & Google Analytics & G72-73 \\
Data as Code & Treat data processing, sharing, and versioning as code. & Data Version Control, Airflow & G74-75 \\
Database as Code & Manage database schemas and configurations as code. & Liquibase, Bytebase, Flyway & G76-77 \\
Docs as Code & Manage documentation using the same tools and processes as code. & Markdown & G78-79 \\
Diagrams as Code & Generate diagrams programmatically from code definitions. & Graphviz, Mermaid & G80-81 \\
Architecture as Code & Model system architectures in an executable, code-centric manner. & Structurizr, Archi & G81-82 \\
Environments as Code & Manage entire development/test/production environments through code. & Terraform, Ansible & G61-63 \\
Platform as Code & Define entire platform components (infrastructure, runtimes, services) in code for easy replication. & Crossplane & G59-60 \\
Storage as Code & Programmatically allocate and configure storage resources. & Ceph & G57 \\
Network as Code & Configure network components by defining the desired state in code. & Netmiko & G58 \\
IAM as Code & Define identity and access roles using code. & HashiCorp's Vault & G67 \\
Privacy as Code & Automate privacy checks with code. & N/A & G68 \\
Project as Code & Accelerate project creation and maintenance via code-based templates. & Yeoman & G83 \\
Orchestration as Code & Codify the execution of orchestration components and services. & Salt Project & G84 \\
\bottomrule
\end{tabular*}
\end{table*}

\subsection{Dimension 2: Functionality and Application}

\begin{figure}[b!]
    \centering
    \includegraphics[width=\columnwidth]{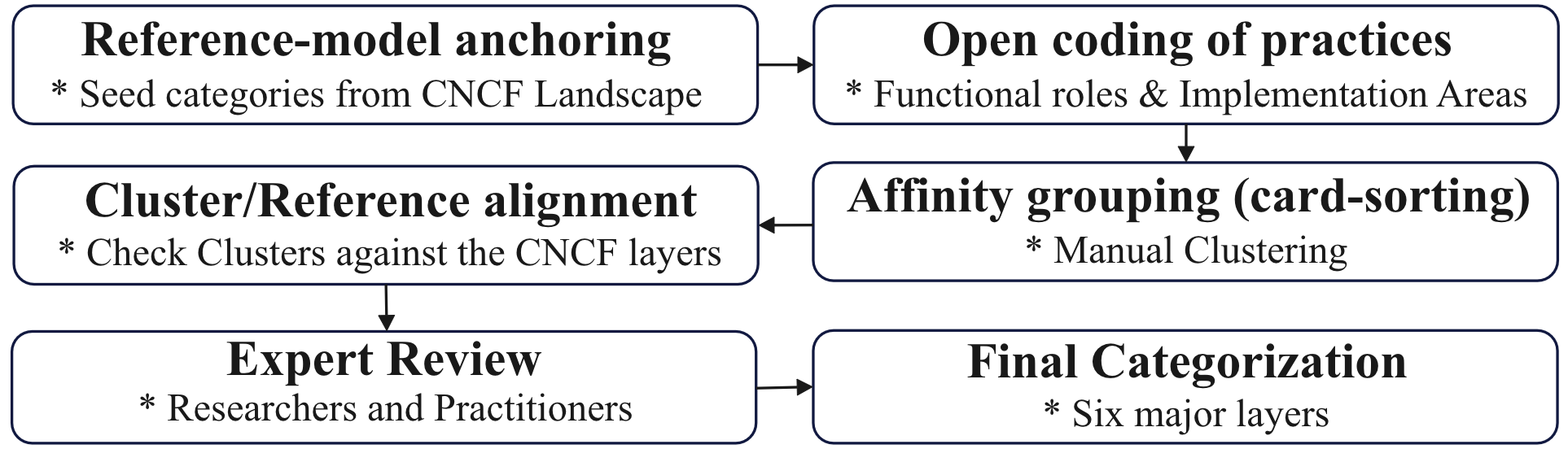}
    \caption{The taxonomy development process: dimension 2.}
    \label{fig:thematic}
\end{figure}

To develop the second dimension of our taxonomy, we adopted the empirical-to-conceptual (bottom-up) method ~\cite{m3} that integrates external evidence with the functional analysis of the identified practices. While the CNCF (Cloud Native Computing Foundation) Landscape \cite{m6} offers a well-established reference for categorizing DevOps technologies, its layers are primarily organized around tooling ecosystems rather than conceptual roles. We therefore used it as a foundational anchor, but conducted an independent thematic analysis~\cite{m2} on the extracted practices to derive a categorization that better reflects the specific scope of EaC (see Fig.~\ref{fig:thematic}). This clustering and refinement resulted in six functional layers, each grouping EaC practices according to their functional roles and application areas, as defined below.

\subsubsection{Infrastructure Provisioning \& Management} Combines CNCF’s \emph{Provisioning} and \emph{Runtime} layers. Practices in this category (e.g., Infrastructure as Code, Storage as Code, Network as Code) declaratively manage foundational resources.

\subsubsection{Platform \& Orchestration} Extends CNCF’s \emph{Orchestration \& Management} and \emph{Platform} layers. It includes practices that define higher-level runtime platforms and coordinate service workloads, such as Platform as Code, Orchestration as Code, and Environments as Code.

\subsubsection{Application Design \& Development} Adapts CNCF’s \emph{App Definition \& Development} layer to include a broader spectrum of application-centric practices. This layer captures code-driven mechanisms for defining application logic and developer workflows, such as Pipeline as Code, Docs as Code, and Architecture as Code.

\subsubsection{Data \& Database} Derived from CNCF’s detailed \emph{Database} layer. Consolidates data-centric practices, such as Data as Code and Database as Code.

\subsubsection{Security \& Compliance} Aligns with CNCF’s detailed \emph{Security \& Compliance} layer. Captures practices dedicated to risk reduction and regulatory assurance, including Policy as Code, Security as Code, and Compliance as Code.

\subsubsection{Observability \& Analysis} Matches CNCF’s \emph{Observability \& Analysis} layer. It includes practices that enable visibility into system behavior and performance through code-managed telemetry, monitoring configurations, and analytic tooling—such as Monitoring as Code, Dashboards as Code, and Analytics as Code.

Beyond assigning practices to these six layers, we also accounted for hierarchical relationships between practices. Specifically, when multiple practices exhibited substantial functional overlap or a clear subset relationship, we consolidated them under a broader umbrella category. For example, \emph{Storage as Code} and \emph{Network as Code} were subsumed into \emph{Infrastructure as Code}, as they all address declarative provisioning of foundational infrastructure components and operate at the same lifecycle stage. 

\begin{figure}[t]
\centerline{\includegraphics[width=\columnwidth,height=\textheight,keepaspectratio]{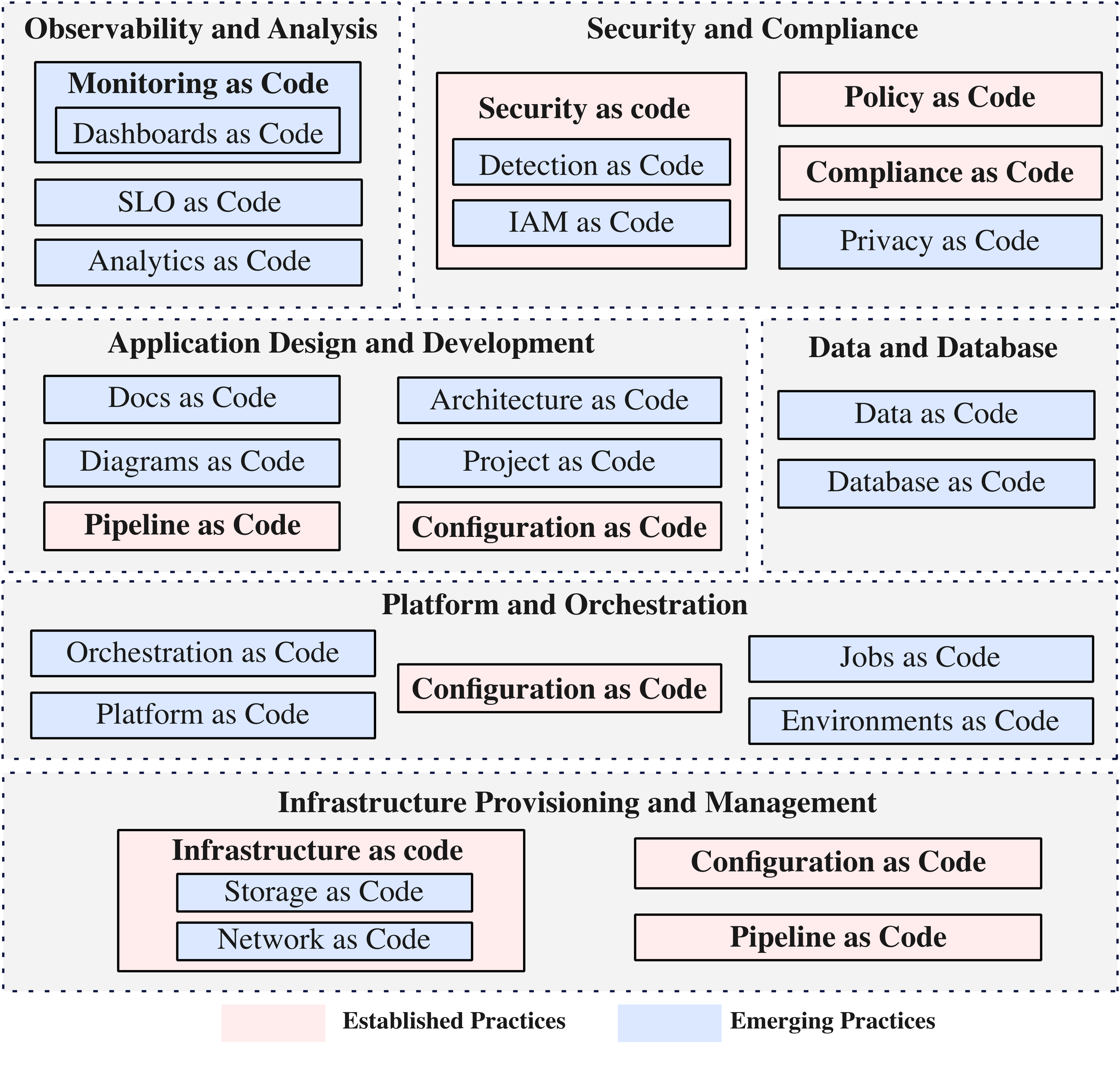}}
\caption{The taxonomy of EaC practices: Functionality and Application.}
\label{fig:taxo_dimension2}
\end{figure}

Additionally, one practice may span multiple layers due to its broad applicability. For example, \emph{Configuration as Code} encompasses post-provisioning configuration of infrastructure (e.g., operating system settings), application configurations (e.g., environment variables), and orchestration platform configurations (e.g., Kubernetes encryption settings). Similarly, \emph{Pipeline as Code} defines CI/CD pipelines as code. While CI/CD pipelines have traditionally been used to automate the build, test, and deployment of applications, they are now also widely applied to automate the testing, provisioning, and management of infrastructure. Fig.~\ref{fig:taxo_dimension2} illustrates the complete categorization, representing the second dimension of our proposed taxonomy.

\section{Conceptual Model}
\begin{figure*}[t]
\centerline{\includegraphics[width=\textwidth,height=\textheight,keepaspectratio]{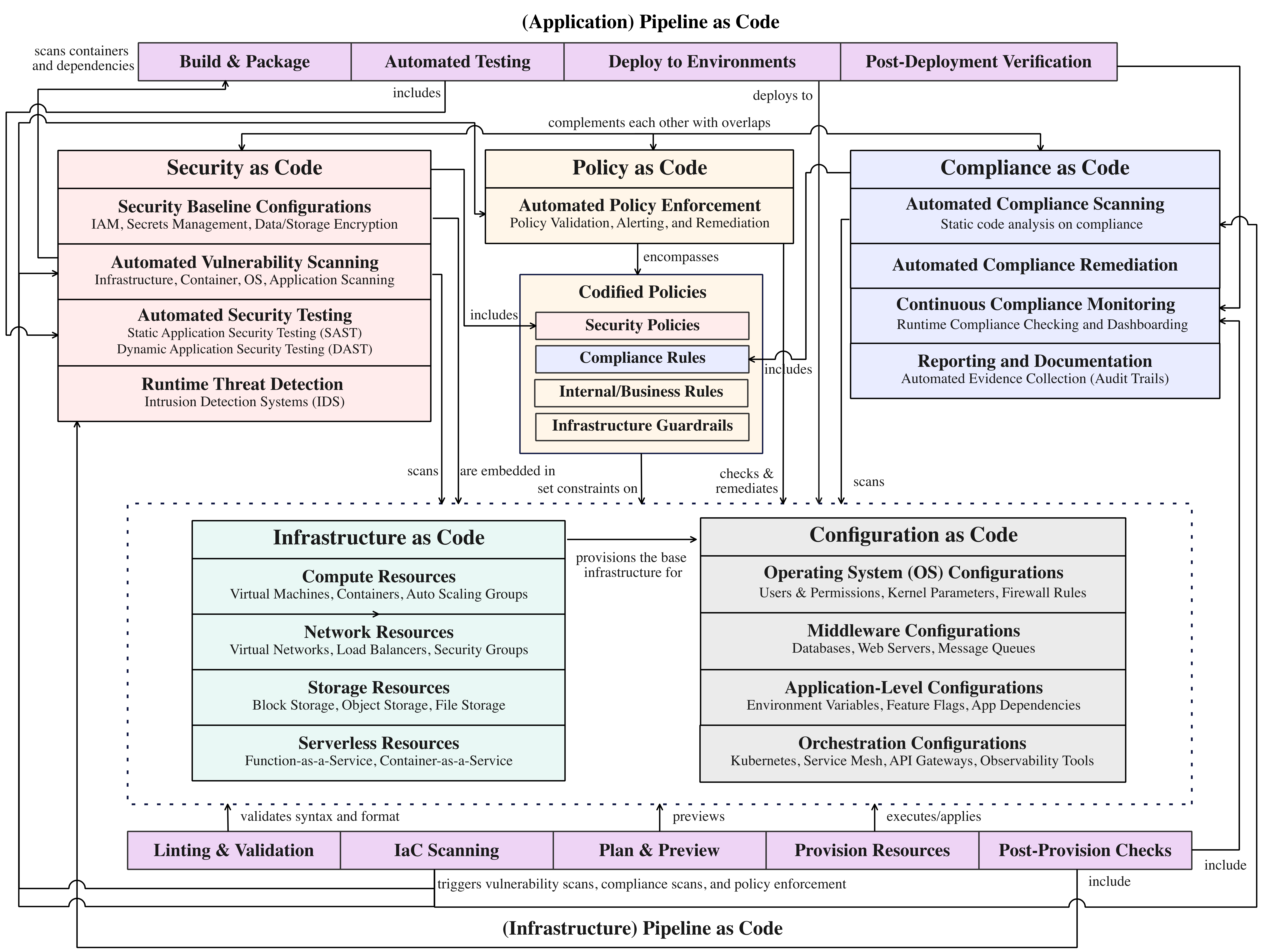}}
\caption{The conceptual model of EaC.}
\label{fig:model}
\end{figure*}

To synthesize a coherent understanding of how EaC practices function and interact, we developed a conceptual model grounded in the data extracted from our literature corpus. This model represents the structural and behavioral relationships between the established practices.

We captured key information about each practice’s definition, scope, functional roles, and relationships (e.g., overlaps, dependencies, integration points) as described in the selected literature sources. The model was constructed through a multi-step process:
\begin{itemize}
    \item \textbf{Subdomain Structuring:} Within each established practice, subdomains were introduced to reflect internal granularity where it was consistently described. For example, CaC was divided into operating system configurations, middleware, application-level settings, and orchestration configurations, based on multiple consistent sources~[G28]-[G31].
    \item \textbf{Concept Mapping:} Extracted relationships were translated into directional connections between practices. Each edge in the model reflects an empirically observed or explicitly stated interaction (e.g., “Policy as Code validates Infrastructure as Code” [G33, G37]).
    \item \textbf{Lifecycle Alignment:} Practices were then mapped to DevOps stages (e.g., build, test, and deploy) based on descriptions of where and how they are applied (see Fig.~\ref{fig:pipeline}). This alignment was guided by recurring patterns in the literature and tool documentation (e.g., static analysis tools applied during testing [G37], runtime compliance tools during operation [G50]).
\end{itemize}
Fig. \ref{fig:model} presents our proposed conceptual model; the details are further described in subsequent subsections.


\subsection{Infrastructure as Code and Configuration as Code}    
Infrastructure as Code (IaC) and Configuration as Code (CaC) are closely related practices. IaC centers on the management and provisioning of infrastructure. CaC, in contrast, focuses on configuring software layers on top of the provisioned infrastructure.

Given these close ties, a natural question arises: why not provision infrastructure and configure its software layer in a single pass using IaC? Tools such as Terraform and Ansible do support both provisioning and post-provisioning tasks. For instance, Terraform’s Provisioners feature enables script or command execution on a resource post-creation (only as a last resort) \cite{d1}. However, relying heavily on these features can create entangled configurations. This undermines the separation of concerns and can complicate version control and state management—key considerations in large-scale production systems \cite{i1}.

In practice, however, the boundary between IaC and CaC can still blur, especially in cloud-native environments. Kubernetes (not traditionally an infrastructure) exemplifies this overlap: it runs on top of virtual or physical machines and provides an infrastructure layer for containerized applications~\cite{d3}. Consequently, IaC tools may provision the underlying infrastructure and simultaneously deploy Kubernetes and its core software components such as kube-proxy (network proxy), CoreDNS (DNS server), and CNI (container network interface) \cite{d4}. This occurs when the two are interdependent. Their dual role—operating as software on infrastructure while also serving as infrastructure for applications—highlights when IaC and CaC can intersect. However, the fine-grained configuration of Kubernetes and its components (e.g., scheduling constraints, auto-scaling settings, ingress controllers) is a CaC task \cite{d5}.
In conclusion, software components are configured with IaC when they are:
\begin{itemize}
    \item Inseparable from the underlying infrastructure.
    \item Treated as core parts of a unified infrastructure unit.
    \item Regarded as foundational to applications, rather than merely installed software.
\end{itemize}
Nevertheless, application-specific configuration (e.g., environment variables, property files, feature flags) falls squarely under CaC \cite{b9}. These settings—whether they define database connection strings, API keys, or feature toggles—are typically managed outside the application codebase \cite{d6}. Doing so preserves flexibility, improves security (by preventing secrets from being hard-coded), and makes it easier to adapt configurations across environments (development, staging, production) without changing the underlying application logic, and independently of infrastructure changes \cite{d6}.

\subsection{Policy as Code, Compliance as Code, and Security as Code}

\begin{figure}[b!]
\centerline{\includegraphics[width=\columnwidth,height=\textheight,keepaspectratio]{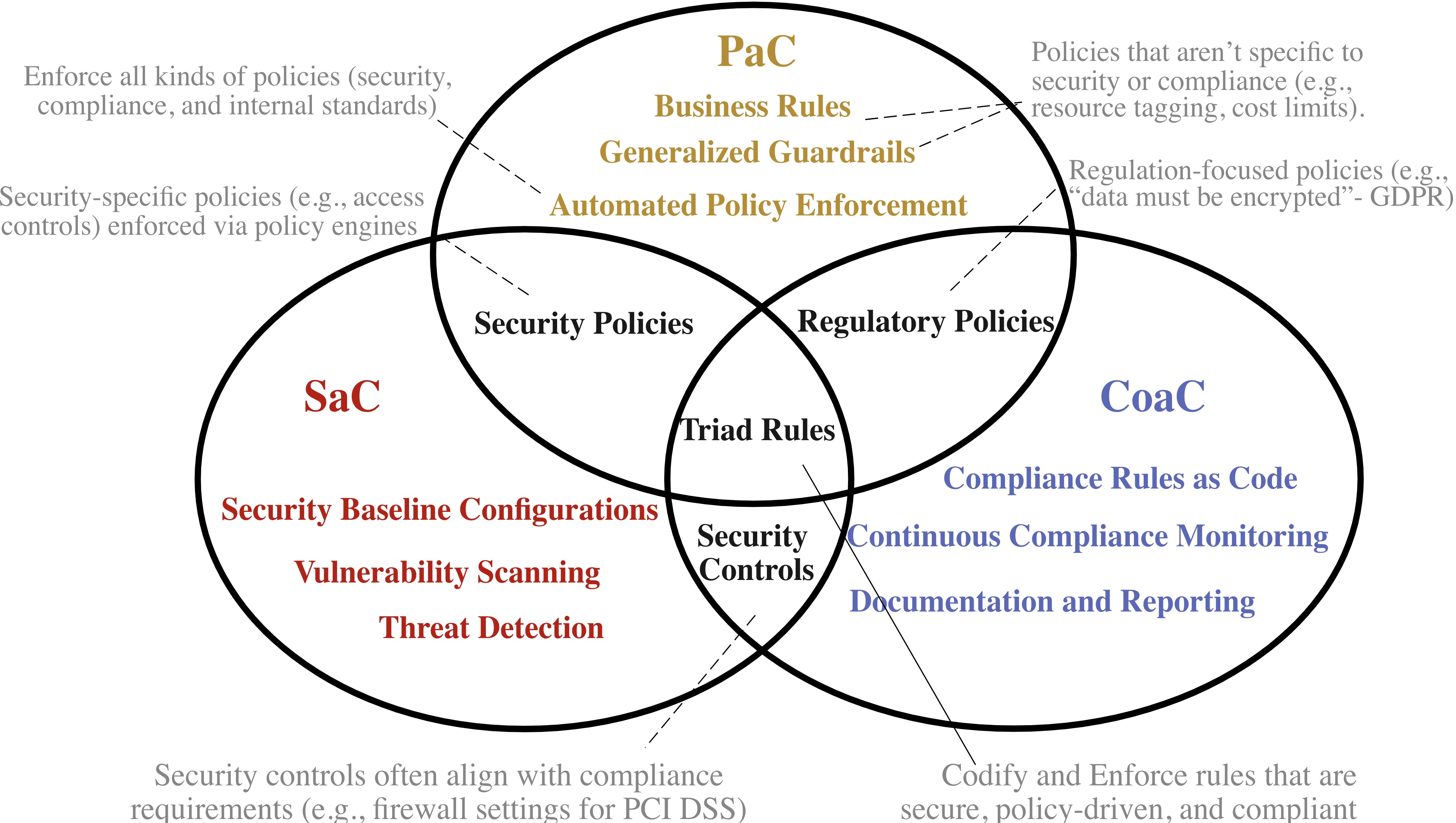}}
\caption{Policy as Code vs. Compliance as Code vs. Security as Code.}
\label{fig:venn}
\end{figure}

In DevOps workflows, security as code, compliance as code, and policy as code converge around their as-code nature—defining rules as executable scripts and automating enforcement through scans, checks, or tool-driven commands. Their boundaries, overlaps, and inter-relations hinge on how these coded implementations are structured and executed. Fig.~\ref{fig:venn} presents a venn diagram showing their overlaps and focuses.

\subsubsection{Policy as Code (PaC)}
Among the three as-code practices, PaC is the most recognized (see Fig. \ref{fig:violin}), with well-defined guidance, dedicated languages (e.g., Rego), and supporting tools (e.g., OPA) \cite{d7}. PaC encompasses a broad range of policies, including security-related rules, regulatory requirements, and internal organizational policies \cite{d17}.

Applications and infrastructures are frequently subject to external regulations (e.g., HIPAA, GDPR, and PCI-DSS) or internal standards (e.g., cost limits, resource tagging, and access permissions) \cite{d19}. Manually verifying adherence to these requirements can be labor-intensive. PaC addresses this challenge by defining machine-readable policies enforced by policy engines, which can block infrastructure provisioning, configuration deployments, or runtime operations if they violate specified rules \cite{d18}. Consequently, PaC’s core objective is governance and enforcement \cite{d17}.

\subsubsection{Compliance as Code (CoaC)}
There is no universally accepted definition of Compliance as Code in academia or industry. In the limited scientific literature, Agarwal et al.~\cite{b11} define CoaC as documenting compliance rules in OSCAL. OSCAL, the Open Security Controls Assessment Language \cite{d16}, is used for documenting regulatory controls (e.g., NIST 800-53 or ISO 27001) and their implementations in machine-readable code (e.g., JSON, XML, YAML). In another work, Yanagawa et al. \cite{b2} demonstrate how CoaC artifacts (OSCAL) can integrate with PaC to validate control implementations. According to their perspective, CoaC focuses on expressing regulatory controls as code, with PaC serving downstream for validation and enforcement.

Industry practitioners diverge in their views of CoaC. Some see it as a subset of PaC centered on enforcing regulatory rules \cite{d14,d15}, whereas others use the term to describe programmatic methods for automating implementation, validation, remediation, monitoring, and reporting of compliance status \cite{d10,d11,d12,d13}.

Bringing these perspectives together highlights substantial overlap between compliance as code and policy as code. From the PaC perspective, compliance policies are one category of a broader set of enforced policies. From the CoaC point of view, the practice can be more comprehensive, leveraging PaC for validation and remediation while also encompassing continuous compliance monitoring and automated reporting of compliance status.

\subsubsection{Security as Code}
Security as Code (SaC) focuses on embedding security practices into automated, code-centric workflows. SaC and PaC have substantial overlap: security policies are one subset of the broader policy set that PaC can enforce such as preventing deployment of container images with known CVEs\cite{d21}—or verifying that Infrastructure or application configurations meet baseline security standards (e.g., verifying encryption at rest or in transit)\cite{d22}.

However, SaC also addresses unique security concerns beyond policy enforcement alone. Examples include:
\begin{itemize}
\item Vulnerability Scanning: Automatically detecting misconfigurations in infrastructure, containers, and OS (e.g., open firewall ports, unsafe IAM permissions) pre-deployment \cite{d27}.
\item Application Security Testing: Embedding static application security testing (SAST)\cite{d23}, dynamic application security testing (DAST) \cite{d24}, and software composition analysis (SCA) \cite{d25} into the CI/CD pipeline as code-driven steps.
\item Threat Detection: Continuously monitoring applications in production for suspicious activities, often underpinned by code-based rules (e.g., threat signatures and anomaly thresholds) \cite{d26}.    
\end{itemize}

Compared to CoaC, SaC is less concerned with mapping specific regulatory controls into machine-readable documentation. Rather, it focuses on automating security best practices. Still, it often intersects with CoaC when those best practices are mandated by regulations (e.g., restricting unauthorized access to meet PCI-DSS requirements). In such cases, SaC can supply enforcement mechanisms and evidence gathering (e.g., vulnerability scan reports) that feed into compliance workflows, while CoaC ensures those controls are traceable to overarching regulations and documented accordingly.

\subsubsection{Analysis}
In summary, the inter-relation stems from their common as-code foundation: rules defined in declarative or imperative formats and enforced via automated scans or tool invocations. A single system component—such as a cloud storage bucket requiring encryption—can be scanned by security tools for threats, checked by compliance scripts for regulations (GDPR), and validated by policy engines for organizational alignment, maximizing efficiency. Yet, this synergy demands precision: security as code prioritizes dynamic scans, compliance as code needs rigid rule-matching, and policy as code requires consistent enforcement. 

\subsection{Pipeline as Code}
Pipeline as Code defines and manages CI/CD pipeline configurations through code, enhancing automation and reproducibility. Traditional CI/CD primarily manages application-specific tasks, including build, testing, and deployment, to facilitate efficient software delivery \cite{d35}. However, infrastructure provisioning—creating and configuring the environments where applications run—is often overlooked. Due to the complexity and dynamic nature of modern, cloud-based infrastructure, maintaining separate infrastructure-specific CI/CD pipelines has become necessary \cite{d36}. Such pipelines handle tasks such as validation of infrastructure configurations (e.g., linting and security scanning), previewing and applying resource changes, and executing post-provision checks (e.g., drift and threat detection) \cite{d37}. 

\begin{figure*}[t!]
\centerline{\includegraphics[width=\textwidth,height=\textheight,keepaspectratio]{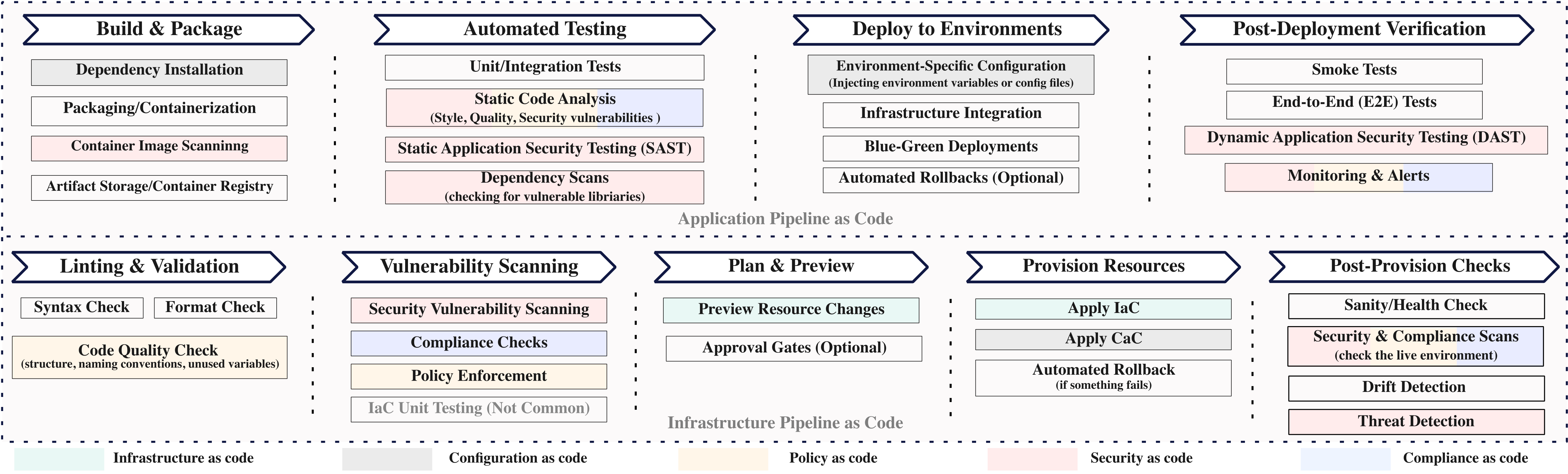}}
\caption{Overview of application and infrastructure Pipelines as Code, highlighting the integration of as-code practices within CI/CD stages.}
\label{fig:pipeline}
\end{figure*}

The application and infrastructure pipelines often integrate other EaC practices for greater consistency and compliance. For instance, Security as Code practices (e.g., SAST and DAST) are embedded in the testing stage of application CI/CD pipelines. Similarly, infrastructure pipelines employ Policy as Code to validate configurations against regulatory rules and organizational standards prior to provisioning. Fig.~\ref{fig:pipeline} illustrates these pipelines, highlighting how various code practices are integrated into the CI/CD stages.

\section{Validation}
To ensure the accuracy, relevance, and applicability of the proposed taxonomy and conceptual model, a rigorous validation process was undertaken involving both academic and industry expertise. We employed an expert-based content validation methodology, integrating checklist-based evaluation and iterative consensus-building to enhance validity.

\subsection{Expert-based Validation and Refinement Process}
We developed two structured validation checklists (for the taxonomy and conceptual model) containing detailed criteria covering:
\begin{itemize}
    \item Content validity (alignment with the literature, clarity of definitions, scope coverage)
    \item Logical consistency (internal coherence, inter-category relationships, interaction representation)
    \item Practical relevance (industry relevance, real-world applicability, scalability) 
\end{itemize}

Three domain experts from our industry partner (a leading cloud service provider)—specifically, a systems architect responsible for high-level infrastructure design, a senior DevOps engineer specializing in automation pipelines, and a senior software engineer with deep experience in cloud-native development—participated in the first validation stage. Each expert independently applied the detailed checklist to every taxonomy and model criterion, using a three-point scale (Yes/Partial/No). They also provided free-form comments to justify their ratings, highlight ambiguities, and suggest refinements.

After collecting these ratings and qualitative annotations, the core author team (two academic researchers and one industry practitioner) conducted a structured review workshop. We began by quantifying inter-rater agreement to identify criteria with low consensus. For each of these items—and for any criteria with substantive qualitative feedback—we held in-depth discussions. During these sessions, we examined the experts’ comments side by side, debated alternative phrasings, merged or split categories where overlap was noted, and clarified relationship definitions. Each change was documented and re-evaluated in subsequent review cycles until all concerns were resolved.

In the second stage, we reconvened all participants in a facilitated consensus meeting. Using the nominal group technique \cite{d34}, we systematically walked through each revised checklist item, presenting our proposed adjustments alongside the original expert feedback. Participants cast “approve/revise” votes on each change, and any item failing to reach unanimous approval was discussed further and modified on the spot. By the end of the session, every checklist criterion—and therefore every element of the taxonomy and model—had been explicitly ratified. The complete validation checklists and results are available online (see Section \ref{sec:resources}).


\begin{table*}[t]
  \centering
  \caption{Summary of Key Expert Comments, Discussions, and Refinements}
  \label{tab:validation_summary}
  \small
  \begin{tabular*}{\textwidth}{@{\extracolsep{\fill}}p{0.12\textwidth}p{0.26\textwidth}p{0.27\textwidth}p{0.28\textwidth}}
    \toprule
    \textbf{Criteria} & \textbf{Comment} & \textbf{Discussion} & \textbf{Refinement} \\
    \midrule
    \multicolumn{4}{c}{\textbf{Taxonomy Validation}} \\
    \midrule
    Scope \& Completeness & 
    Certain ``as-code'' practices (e.g., Law as Code, Contracts as Code, Management as Code) were not included. & 
    These practices currently lack sufficient literature support, widespread recognition, or clear relevance to typical software engineering workflows. & 
    Created supplementary files documenting these speculative practices for future consideration. \\
    \midrule
    Scalability for New Practices & 
    Unclear how new practices could easily be integrated into the taxonomy. & 
    Taxonomies inherently cannot anticipate all emerging practices, but flexible representation is beneficial. & 
    Converted taxonomy into a machine-readable JSON format to facilitate easier future additions and modifications. \\
    \midrule
    Granularity \& Inter-categorical Relationships & 
    Categories could be insufficiently granular; inter-categorical relationships were not clearly described. & 
    Categories and relationships were intentionally aligned with CNCF Landscape, an established industry model. & 
    Provided citations and detailed mappings between taxonomy categories and CNCF layers for clarity (Section IV-B). \\
    \midrule
    \multicolumn{4}{c}{\textbf{Conceptual Model Validation}} \\
    \midrule
    Interaction Granularity & 
    Relationships in the conceptual model were recommended to reflect greater detail and granularity. & 
    Earlier detailed iterations proved overly complex and compromised readability. & 
    Produced a companion document explicitly detailing each relationship to preserve main model clarity. \\
    \midrule
    Tool Integration \& Practical Application & 
    Practical integration with industry-standard tools was not obvious for less-experienced developers. & 
    Model aims primarily at conceptual clarity, yet practical applicability remains crucial. & 
    Developed Implementation Examples providing concrete code snippets demonstrating model integration with industry-standard tools. \\
    \midrule
    Model Scalability & 
    Limited evidence indicating ease of extending the model to include emerging practices and relationships. & 
    Extending the model requires significant domain expertise and deep structural understanding, making scalability challenging. & 
    Proposed future conversion into machine-readable format, with AI-assisted updates to enhance adaptability and scalability. \\
    \bottomrule
  \end{tabular*}
\end{table*}

\subsection{Key Comments and Refinements}
Expert validation identified several areas for improvement regarding both our taxonomy and conceptual model, along with suggestions and acknowledged limitations (summarized in Table~\ref{tab:validation_summary}). The taxonomy received feedback primarily concerning its completeness, scalability for future practices, and granularity of categorization and relationships. To address these comments, we have created supplementary resources for emerging practices, implemented a machine-readable JSON format to enhance scalability, and clearly aligned our categories with the CNCF Landscape.

The conceptual model similarly prompted feedback focused on interaction granularity, practical integration with industry-standard tools, and overall scalability. To refine the model, we introduced an extra relationship documentation, practical code examples, and proposed future solutions for extensibility and scalability.

\subsection{Limitations}
The taxonomy and conceptual model face limitations in terms of time-sensitivity and scalability. Since the taxonomy is based on past literature, it is inherently time-sensitive and must be easy to scale and update as technology evolves. However, modifying the model is currently a manual process that requires deep domain expertise in their functionality and integration with existing practices. To address this, we propose future work leveraging AI-assisted augmentation to automate the classification and integration of new EaC practices. For example, large language models (LLMs), fine-tuned on a curated dataset of EaC literature, can be employed to suggest initial categorizations and integration pathways for new practices.

Beyond these conceptual challenges, the actionable guidance we provide for practitioners to adopt our taxonomy and model remains insufficient. While the current code snippets (Section~\ref{sec:resources}) offer preliminary examples of basic implementation, they lack the depth and contextual detail required to address real-world complexities. To bridge this gap, comprehensive case studies should be conducted to illustrate how to overcome organizational barriers and adapt the taxonomy to diverse environments effectively.

\subsection{Threats to Validity}
Despite our efforts to ensure a thorough validation process, threats to validity persist. First, the expert-based validation relied on a limited panel of three industry experts from a cloud service provider, which may introduce bias and limit the generalizability of feedback across diverse organizational contexts. Second, the subjective nature of expert evaluations, even with structured checklists and consensus-building, may still reflect individual perspectives rather than universal truths, potentially affecting the objectivity of the validation outcomes.

\section{Supplementary Resources}\label{sec:resources}
To enhance transparency, reproducibility, and practical utility, we provide several supplementary resources \cite{d38} associated with this study. These materials are intended to support readers in further exploring the taxonomy and conceptual model, as well as to facilitate future research and adaptation:

\begin{itemize}
  \item \textbf{Selected Literature and Extracted Data:} A list of scientific and grey literature documents reviewed in the study, along with the data extracted from them.
  \item \textbf{Validation Checklist and Results:} Expert evaluation criteria, ratings, agreement metrics, and comments.
  \item \textbf{Extended Taxonomy:} A more comprehensive version of the taxonomy with detailed descriptions.
  \item \textbf{Conceptual Model Companion File:} A detailed specification of the components and relationships within the conceptual model.
  \item \textbf{Code Examples:} A curated set of annotated code snippets using industry-standard tools, demonstrating the practical implementation of the conceptual model.
\end{itemize}

\section{Conclusion and Future Work}

This paper introduced a comprehensive taxonomy and conceptual model for Everything as Code (EaC), developed through a multivocal literature review, quantitative analysis, and thematic synthesis. The taxonomy classifies 25 EaC practices based on their industry recognition and functional roles. Complementing this, the conceptual model illustrates how these practices interact and integrate across the software delivery lifecycle.

Collectively, these contributions address the current scarcity of academic discourse on EaC by offering a structured overview of its scope, boundaries, and interrelationships. They provide improved conceptual clarity, actionable guidance for practitioners, and a foundation for future academic research in this domain.

To validate the artifacts, we employed an expert-based methodology involving both researchers and industry practitioners. The validation process, which included structured evaluation and iterative consensus-building, affirmed the accuracy, relevance, and practical applicability of the taxonomy and model while also revealing areas for improvement.

Future work will address these limitations, particularly concerning the incorporation of emerging EaC practices and scalability of the framework. We plan to encode both the taxonomy and conceptual model in machine-readable formats and explore AI-assisted augmentation—such as leveraging large language models (LLMs) to interpret and recommend the placement of new practices—thereby supporting automated updates and ensuring long-term adaptability.

\section*{Acknowledgment}
This research was supported in part by IBM Canada
Center for Advanced Studies (CAS) and the Natural Sciences and Engineering Research Council of Canada (NSERC).

\end{document}